\documentclass[doublespacing]{elsart}
\usepackage{graphicx}  
\usepackage{amssymb} 
\usepackage{color}

\def\be{\begin{eqnarray}}
\def\ee{\end{eqnarray}}

\def\3half{{\textstyle \frac32}}

\def\la{{\langle}}
\def\ra{{\rangle}}

\begin{document}
\begin{frontmatter}

\title{  Parity non-conservation in nuclear excitation by
circularly polarized photon beam
}

\author{A.I.Titov$^{1,2}$, M. Fujiwara$^{1,3}$ and K. Kawase$^{3}$ }

\address{
$^1$ Kansai Photon Science Institute, Japan Atomic Energy
              Agency, Kizu, Kyoto 619-0215, Japan \\
$^{2}$Bogoliubov Laboratory of Theoretical Physics, JINR,
               Dubna 141980, Russia
$^3$Research Center for Nuclear Physics, Osaka University,
Ibaraki, Osaka 567-0047, Japan
}

\begin{abstract}
We show that the asymmetries in the nuclear resonance
fluorescence
processes with a circular polarized photon beam
 may be used as a tool for studying  the parity
non-conservation (PNC) in nuclei. The PNC asymmetry
measurements both in exciting the parity doublet states and in exciting
the discrete states near the ground states with parity mixing
are discussed. We derived the formulae needed for measuring
the PNC asymmetries.

\begin{flushleft}
\end{flushleft}
\end{abstract}
\end{frontmatter}



Parity non-conservation (PNC) is well known after the
discovery of the mirror symmetry  violation in ${\beta}$-decays by Wu
\cite{Wu}, following the suggestion by Lee and Yang
\cite{Lee}.
The origin of this mirror-asymmetry is now clearly
understood as a manifestation of the exchange processes of weak bosons,
$W^{\pm}$, which are mediators of ${\beta}$-decay.

Observations of the  PNC effect in the
nucleon-nucleon interaction are not new.
The trial to observe the PNC effect began with the first report  by
Tanner in 1957 \cite{Tanner}, followed by the famous work of
Feynman and Gell-Mann \cite{F-G} for the universal current-current
theory of weak interaction.
Wilkinson \cite{Wilkinson} also triggered the enthusiastic
long studies of finding the tiny PNC effect
in  nuclear excitation processes.

The process contributing to the PNC effect is due to the non-trivial quark
interactions with  weak $Z^0$ and $W^\pm$  bosons  in the effective
nucleon-nucleon-meson vertices.
The details of the PNC studies were reviewed in
Refs. \cite{AH85,Desp98}. However, the current problem is
essentially focused on the fact that the weak meson-nucleon
coupling constants (and in particular, the pion-nucleon
coupling constant) deduced from various  experiments
are not consistent.
It is concluded by Haxton {\it et al.,} \cite{Haxton}
that the experimental PNC studies are
still  not satisfactory, and more experimental as well as
theoretical studies are needed.

One of  experimental PNC studies is  to measure the
parity mixing  between the parity-doublet states.
On the basis of the first order perturbation
theory,  the wave functions of two closely-located
states, ${\mid}{{\phi}_{\pi}}{\rangle}$ and
${\mid}{{\phi}_{-\pi}}{\rangle}$ are mixed by the PNC
interaction $V_{PNC}$ as
\begin{equation}
 {\mid}\tilde{{\phi}_{\pi}}{\rangle} =
{\mid}{\phi}_{\pi}{\rangle}  +
 \frac{{\langle}{\phi}_{-\pi} {\mid} V_{\rm PNC} {\mid}
 {\phi}_{\pi}{\rangle}}{E_{\pi}-E_{-\pi}}
 {\mid}{\phi}_{-\pi}{\rangle},
 \label{E1}
\end{equation}
where  $\phi_{\pi}$ and $E_{\pi}$ are the wave function
and the excitation energy
of the levels with the same spin and opposite parity
($\pi=\pm$), respectively.
In the traditional experiment, one of the doublet levels
is excited  in a nuclear reaction.
Then, the PNC effect appears in the
asymmetry of emitted circularly polarized
photons~\cite{AH85}.
 \begin{equation}
 P_{\gamma}\sim \frac{2\,R}{E_{\pi}-E_{-\pi}}
{\langle}{\phi_{-\pi}}
{\mid}
 V_{\rm PNC} {\mid} {\phi}_{\pi}{\rangle}~,
 \label{E2}
 \end{equation}
where $R\gg1$ is the ratio of the nuclear electromagnetic
transitions with opposite parities.
Thus, for example, in
the case of~~$^{21}$Ne, the level distance between
the two excited
1/2$^{-}$ and 1/2$^{+}$ states at 2789 keV and 2795 keV is
only 5.7 keV, and a mixing of the order of about
$300{\times}10^{-4}$
is expected. One experiment for  $^{21}$Ne was performed
by the Seattle group~\cite{Earle83} and  the asymmetry of  $(0.8
\pm 1.4) \times 10^{-3}$ was reported.
Even though the nuclear
amplifier factor is rather large,
the accuracy of the measurement
gives only upper bound of the nuclear PNC-effect.
Other examples are
reported in the E1 and M1 mixing transitions for $^{19}$F(1/2$^-$,
109.9 keV ${\rightarrow}$ 1/2$^+$, g.s.), $^{18}$F(0$^-$, 1.080
MeV ${\rightarrow}$ 1$^+$, g.s.), and $^{175}$Lu (9/2$^-$, 396
keV ${\rightarrow}$ 7/2$^+$, g.s.) (for details, see Refs.
\cite{AH85,Desp98,Holstein,HH95}).
In all cases, the prime
problems to be solved are insufficient data accuracy and
its interpretation for the transition matrix using a model
calculation.
Obviously, it is still important to obtain different
types of reliable experimental results for checking the
difficult PNC measurement.

In this paper, we wish to point out a possibility to use a
circular polarized high intensive ${\gamma}$-ray beam for
the PNC studies via the nuclear resonance fluorescence (NRF)
processes.
Assuming a circular polarized ${\gamma}$-ray beam is
intense, one can  excite the parity doublet states and observe the
de-excitation of the ${\gamma}$-rays.
The parity non-conservation will appear in the difference
of the photon absorption with
different helicities of the incoming ${\gamma}$-rays.

In fact, the asymmetry of the photon absorption $A^a_{RL}$
\begin{eqnarray}
A^a_{RL}=\frac{\sigma^a_R-\sigma^a_L}{\sigma^a_R+\sigma^a_L},
\end{eqnarray}
where $\sigma^a_{R(L)}$ stands for the photon absorption
cross sections with the right (left)-circularly polarized
photons is equal to the asymmetry $P_\gamma$ in Eq.~(\ref{E2}) which
was discussed in many papers \cite{AH85,Desp98,Holstein,HH95}.
The new aspect discussed here is
that the asymmetry in reaction $\gamma_i+ A_{gs}\to A^* \to
\gamma_f+A_{gs}$, $A_{RL} (\theta)$, is, in general, different from
$A^a_{RL}$.
The PNC asymmetry in the NRF process depends on the
the angle ($\theta$) between the directions of flight of
absorbed and emitted photons, and this dependence can enhance or
reduce the PNC-effect.

Let us consider electromagnetic excitation and decay of
the lowest excited $\frac{1}{2}^-$ ($E_x=109.9$ keV) state in
$^{19}$F.
This example is very transparent and can be easily extended to
other parity doublets  with higher spins.
Therefore, we use it as
a starting point of our consideration, leaving discussion of
practical utilization of this particular transition, which
may be not so easy at once because of the finite life time of the
radiative $^{19}$F.
It is assumed that the ground state with
$J^{\pi}=\frac12^+$ and the first excited state with
$\frac12^-$
are the parity doublet
 \begin{eqnarray}
 && |\widetilde{\frac12^+}\ra \simeq | \frac12^+\ra -
\alpha
|\frac12^- \ra,
 \nonumber\\
 && |\widetilde{\frac12^-}\ra \simeq | \frac12^-\ra +
\alpha
|\frac12^+ \ra,
 \label{E3}
 \end{eqnarray}
with
 \begin{eqnarray}
 \alpha={ \la \frac12^-|V_{\rm PNC}|\frac12^+\ra }/{\Delta
E},
 \label{E4}
 \end{eqnarray}
and $ {\Delta E}=E_{\frac12^-}-E_{\frac12^+}$.
The amplitude of
the process $\gamma_i+A\to A^*\to \gamma_f+A$ ($A=^{19}$F)
may be
expressed as a product of absorption ($T^a$) and decay
($T^d$)-
amplitudes
\begin{eqnarray}
 T_{\lambda_i\lambda_f}=T^a_{m*;\lambda_i,m_i}\cdot
 T^d_{\lambda_f,m_f;m^*},
 \label{E5}
 \end{eqnarray}
where $m_i,m^*$, $m_f$, and $\lambda_i$, $\lambda_f$ are
the spin projections of the nucleus $A$ in the initial, excited,
and the final states,
and the photon helicities in the initial and
the final states, respectively.
Here, we assume that the spin projection
 of the excited state is conserved during its short decay
 lifetime.
 \begin{figure}[ht] \centering
\includegraphics[width=.35 \columnwidth]{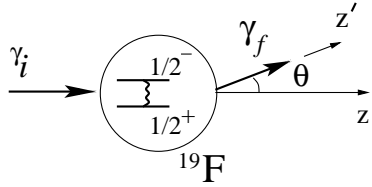}
 \caption{\label{fig:1}
{\small Reaction scheme of $\gamma_i$ + $^{19}$F
($\frac{1}{2}^+$)
 $\to$ $^{19}$F$^*$ ($\frac{1}{2}^-$)   $\to$ $\gamma_f$ +
$^{19}$F ($\frac{1}{2}^+$) } }
 \end{figure}

First, consider the  absorption of the circularly
polarized photon.
The general form of the nuclear electromagnetic transition
amplitude  in the obvious standard notations
reads~\cite{Walecka}
\begin{eqnarray}
 T_{J_f m_f;\lambda, J_i m_i}  &=& - \sum_{L \ge 1}i^L
\sqrt{2\pi (2L+1)} \times \frac{\langle J_i m_i L
\lambda|J_fm_f\rangle}{\sqrt{2J_f+1}} [F_{EL} + \lambda
F_{ML}],
 \label{E6}
 \end{eqnarray}
 where $J_{i,f}$ and  $m_{i,f}$ are the spin and spin
projection  of the initial and the final states, $\lambda$ is the
photon helicity,  $F_{E/M\,L}=\la f || T^L_{E/M}||i\ra $
 is the reduced matrix element of the multipole
 operators.
 In the case of $J_i=J_f=\frac{1}{2}$,  we have
\begin{eqnarray}
 T_{m_f;\lambda, m_i}  =
 i(\lambda E1 +M1)\delta_{m_f,\frac{\lambda}{2}}\delta_{m_i,-\frac{\lambda}{2}},
 \label{E71}
 \end{eqnarray}
 where we denote $E1\equiv \sqrt{2\pi}F_{E1}$ and
$M1\equiv\sqrt{2\pi}F_{M1}$.
 The amplitude of the electromagnetic transition in the
parity doublet
 of Eq.~(\ref{E3}) reads
\begin{eqnarray}
 T_{m^*;\lambda_i, m_i}({\bf z})
 =i\left(\lambda_i E1 + \alpha\mu\right)
\delta_{m^*,\frac{\lambda_i}{2}}\delta_{m_i,-\frac{\lambda_i}{2}}~,
 \label{E8}
 \end{eqnarray}
where $E1$ is the dipole electric transition, and $\mu$ is
the
difference of magnetic moments of the ground $(\mu^+)$ and
excited
$(\mu^-)$  states, respectively: $\mu=\mu^+ - \mu^-$, with
\begin{eqnarray}
 \mu^\pm &=& \sqrt{2\pi}\la
\frac12^\pm||T^{1}_M||\frac12^\pm\ra.
 \label{E72}
 \end{eqnarray}
In Eq.~{\ref{E71}} we stress the dependence of the
transition amplitude on the quantization axis {$\bf z$}, explicitly.
In our convention for the transition amplitude, the quantization axis
coincides with the direction of the incident photon momentum.

Let us turn to the decay of the excited state $^{19}$F$^*$
$\to$
$\gamma_f$ + $^{19}$F.
 The corresponding  amplitude reads
\begin{eqnarray}
 T_{\lambda_f, m_f;\bar m^*}({\bf z'})
 =-i\left(\lambda_f E1 + \alpha\mu \right)^*
 \delta_{\bar m^*,\frac{\lambda_f}{2}}\delta_{m_f,-\frac{\lambda_f}{2}}.
 \label{E9}
 \end{eqnarray}
The main difference between absorption and decay
amplitudes comes from the
 difference in the quantization axis. Now it is
fixed along the direction of flight of the outgoing
photon.
The spin polarizations of the exited state in the two frames
are related to each other as
\begin{eqnarray}
 |\frac12,\bar m'\ra=d^{\frac12}_{ mm'}(\theta)|\,\frac12,m\ra,
 \label{E10}
 \end{eqnarray}
where $d^j_{mm'}(\theta)$ is the Wigner function, and
$\theta$ is
the angle between the beam direction and the direction of
flight of the emitted photon. This relation leads to
\begin{eqnarray}
 T_{\lambda_f, m_f;\frac{\lambda_f}{2}}
 =-i\left(\lambda_f E1^* + \alpha(\mu)^*\right)
d^{\frac12}_{\frac{\lambda_i}{2}\frac{\lambda_f}{2}}(\theta)
 \delta_{m_f,-\frac{\lambda_f}{2}}~.
 \label{E11}
 \end{eqnarray}
 Taking into account
\begin{eqnarray}
 \sum\limits_{\lambda_f}(d^{\frac12}_{\frac{\lambda_i}{2}\frac{\lambda_f}{2}}(\theta))^2
 &=&1, \nonumber\\
 \sum\limits_{\lambda_f}
\lambda_f(d^{\frac12}_{\frac{\lambda_i}{2}\frac{\lambda_f}{2}}(\theta))^2
 &=&\lambda_i\cos\theta,
 \label{E12}
 \end{eqnarray}
 and neglecting the terms proportional to $\alpha^2$,
 we get the PNC-asymmetry in the following form
\begin{eqnarray}
  A_{RL}(\theta) =(1+\cos\theta)< A_{RL}>~,
 \label{E13}
\end{eqnarray}
with
\begin{eqnarray}
  <A_{RL}>  =
  2\alpha  {\rm Re}\left(\frac{\mu}{E1}\right),
 \label{E14}
\end{eqnarray}
One can see that the PNC-asymmetry in the reaction
$\gamma_i+^{19}$F $(\frac12^+) \to ^{19}$F$(\frac12^-) \to
\gamma_i+^{19}$F $(\frac12^+)$ has ``$1+\cos\theta$''-dependence.
It is enhanced (suppressed) at $\theta\sim 0$ ($\theta\sim\pi$).

This idea can be extended to the other nuclei.
In case of
$^{18}$F, there are parity doublet states with $J^\pi=0^+$
and $0^-$ at the energies $E_x=1.042$ and 1.081 MeV,
respectively.
Although the $1^+$ ground state of $^{18}$F is unstable
and thus
the experimental feasibility is very low, we show the
result of
corresponding calculation as a prediction for any $1^+\to
0^-(0)^+$ transitions.

 In case of the transition $\gamma +(1^+)\to (0^-)$[1081
   keV]$\to\gamma + (1^+)$ in $^{18}$F, the asymmetry is
isotropic,  because the excited state with $J=0$ loses
information
about the spin-helicity in the initial state
\begin{eqnarray}
 A_{RL}(\theta)= <A_{RL}>=2   \frac{\la 0^- |V_{\rm
PNC}|0^+\ra
 }{E_{0^-}-E_{0^+}}
 {\rm Re}\left(\frac{M1}{E1}\right).
 \label{E15}
\end{eqnarray}
Here $E1$ and $M1$ are the amplitudes of the $1^+\to 0^-$
and
$1^+\to 0^+$ transitions, respectively.

The PNC asymmetry for the transition  $\gamma
+(\frac32^+)\to(\frac12^-)$[2789
keV]$\to\gamma + (\frac32^+)$ in ${}^{21}$Ne has the
following form
\begin{eqnarray}
 A_{RL}(\theta)  \simeq   (1 +
 \frac14{\cos\theta} )<A_{RL}>,
 \label{E16}
\end{eqnarray}
where
\begin{eqnarray}
 <A_{RL}>  =  -2 \frac{ \la \frac12^+|V_{\rm
PNC}|\frac12^-\ra }
 {E_{{\frac12}^+}-E_{{\frac12}^-}}
 {\rm Re}\left(\frac{M1}{E1}\right),
 \label{E17}
\end{eqnarray}
where  $E1$ and $M1$ are the amplitudes of the
$\frac{3}{2}^+\to\frac{1}{2}^-$ and $\frac{3}{2}^+\to \frac{1}{2}^+$
transitions,
respectively,
and the terms proportional to $M2/E1$ and
$E2/M1$ are neglected. The factor $\frac14$ in Eq.~(\ref{E16})
reflects the fact that the spin projection $m_{i}$ of the ground
state at the fixed photon helicity $\lambda_i$ may be
$-\frac12\lambda_i$ or $-\frac32\lambda_i$.


Instead of the reactions with polarized photons and
unpolarized target,
one can analyze the reactions with polarized target and
unpolarized beam.
The spin asymmetry is defined as
\begin{eqnarray}
  A_{S} =\frac{\sigma_+ - \sigma_-}{\sigma_+ + \sigma_-},
 \label{E18}
\end{eqnarray}
where $\sigma_{\pm}$ stands for the cross section with the
target
polarization $M_i/J_i=\pm 1$, and the quantization axis is
along the beam direction. The corresponding asymmetries are
related to the photon asymmetries $A_{RL}$ as following:\\
$^{19}$F
\begin{eqnarray}
 A_{S}(\theta)=-A_{RL}(\theta)= -  (1 +\cos\theta
)<A_{RL}>~;
 \label{E19}
\end{eqnarray}
$^{18}$F
\begin{eqnarray}
 A_{S}(\theta)=-A_{RL}(\theta)=-<A_{RL}>~;
 \label{E20}
\end{eqnarray}
and $^{21}$Ne
\begin{eqnarray}
 A_{S}(\theta)= -  (1 +\frac{1}{2}\cos\theta )<A_{RL}>~.
 \label{E21}
\end{eqnarray}


In Table~\ref{tab:1}, we show other possible examples for
studying
the PNC-transitions in light nuclei by NRF.
For completeness, we also
show the corresponding angular correlations for the photon
asymmetries
for transitions with the spin in initial and final states
are $J_i$ and $J_f$, respectively.\\
Transitions $0 \to 1$
\begin{eqnarray}
 A_{RL}(\theta)=\left(1
+\frac{\cos\theta}{1+\cos^2\theta}\right)
 <A_{RL}>~.
 \label{E22}
\end{eqnarray}
Transitions $0 \to 2$
\begin{eqnarray}
 A_{RL}(\theta)=\left(1
+\frac{2\cos\theta(2\cos^2\theta-1)}{1-3\cos^2\theta
 +4\cos^4\theta}\right)
<A_{RL}>~.
 \label{E23}
\end{eqnarray}
Transitions $1 \to 1$
\begin{eqnarray}
 A_{RL}(\theta)=\left(1
+\frac{2\cos\theta}{5+\cos^2\theta}\right)
 <A_{RL}>~.
 \label{E24}
\end{eqnarray}
Transitions $1 \to 2$
\begin{eqnarray}
 A_{RL}(\theta)=\left(1
+\frac{90\cos\theta}{73+21\cos^2\theta}\right)
 <A_{RL}>~.
 \label{E25}
\end{eqnarray}
In the equations described above, the average value of the
asymmetry is defined
as a the product of the PNC-matrix element and the nuclear
amplifier factor
\begin{eqnarray}
 |<A^i_{RL}>|=2\vert\frac{R_N^i}{\Delta E_i}
\la|V_{\rm PNC}|\ra_i\vert~.
 \label{E26}
\end{eqnarray}


 \begin{table}[t!]
  \caption{Possible candidates for studying the PNC
asymmetry in
  the light nuclei. The energy levels and
  the amplifier factors $|R_N/\Delta E|$ are given in keV
and
  MeV$^{-1}$, respectively.}
  \label{tab:1}
 \begin{tabular}{lrrrll}
\hline\hline
 $^A$Z & ${\rm transition}\atop{(J^\pi_i;I_i)[E_i]\to
  (J^\pi_f;I_f)}$&$[E_f]$
 &${\rm admixture}\atop{(J_f^{-\pi})[E_f']}$ &
$|R_N/\Delta E|$&
 \\ \hline
 $ ^{14}$C&  $(0^+,1)\to (2^-,1)$&$\,[7340]$ &\, $[7010]$ &\,$ 31\pm6  $ &\\
 \hline
$ ^{14}$N&  $(1^+,0)\to (1^+,0)$&$\,[6203]$ & $[5691]$
 &\,$ 7.0\pm2.0  $ &\\
 ~~~~~~~ &  $(1^+,0)\to (0^+,1)$&$\,[8624]$ & $[8776]$
 &\,$ 40\pm 5  $ &\\
 ~~~~~~~ &  $(1^+,0)\to (2^-,1)$&$\,[9509]$ &$[9172]$
 &\,$ 45\pm 5  $&\\
 \hline
$ ^{15}$O&  $(\frac12^-,\frac12)\to (\frac12^-,\frac12)$&$\,[11025]$ &  $[10938]$ &\,$ 37\pm7  $ &\\
 \hline
 $ ^{16}$O&  $(0^+,0)\to (2^-,0)$& $\,[8872]$ &\, $[6917]$
 &\,$ 18\pm 2  $& \\
 ~~~~~~~&~~~~~~~~ &  ~~~~~~~~~ &\, $[11520]$
 &\,$ 9.5\pm  0.7 $&\\
 \hline
$ ^{18}$F&  $(1^+,0)\to (1^-,0+1)$&$\,[5605]$ & $[5603]$ &\,$ 590\pm110 $& \\
 \hline
$ ^{20}$Ne&  $(0^+,0)\to (1^-,0)$&$\,[11270]$ & $[11262]$
 &\,$670\pm7000   $&\\
\hline
\hline
\end{tabular}

 \vspace*{0.5cm}
\end{table}

In summary, we have discussed the possibilities for
studying the PNC asymmetries in the nuclear resonance fluorescence
processes.
This measurement is inverse compared to the widely
discussed processes for the measurements of circular polarized
$\gamma$-rays from the  radioactive sources. In the previous experiments
for the circular polarization measurement of emitted
${\gamma}$-rays,
one of the parity doublet levels was excited via a nuclear
reaction, and the admixture of the configuration of the opposite
parity was manifested as the asymmetry $A_{\gamma}$ of
${\gamma}$-rays emitted from the excited states with a polarization,
or as the circular polarization $P_{\gamma}$ of ${\gamma}$-rays
emitted from unpolarized excited states.
The isospin-structure of the corresponding transitions in
each of
these cases are different. This  results in different
structure of
the transition matrix elements, and therefore it is
possible to
get independent information on the elementary
parity-violated
meson-nucleon coupling constants.

At present, there is no data available to measure the PNC
effect with circular polarized photons although there are
theoretical estimations for the PNC effect in the deuteron
photodisintegration
\cite{FT04,Liu04} for which the $A_{RL}$ asymmetry are
expected to be very small as the $10^{-7}$ level.

In case of the PNC measurement for the
transition from the 1/2$^+$ ground state to the
109.9 keV 1/2$^{-}$ state in $^{19}$F, for
example, a high intensity photon source from the synchrotron
radiation facilities at SPring-8 is useful.
The intensity of photons
from a elliptical multipole wiggler system
at SPring-8 \cite{Mare1998} reaches at around 10$^{13}$ photons/second
even at $E_{\gamma}$=109.9 keV with an energy width (${\Delta}_E$) of
100 eV. The expected yield rate R of the
${\gamma}A{\rightarrow}A^{*}{\rightarrow}{\gamma}A$ reaction reads
\cite{Skorka75}
\begin{eqnarray}
 R  &=&
 {\pi}^{2}{\lambda}^2\frac{\Gamma}{\Delta_E}\,I_i{\rho}\,d\,N_A/A_t,
 \label{E27}
 \end{eqnarray}
where ${\Gamma}=7.7{\times}10^{-7}$ eV is the resonance width
\cite{AJ72}, ${\lambda}={\hbar c}/E_{\gamma}{\simeq}1.79{\times}10^{-10}$ cm,
d and ${\rho}$ are the target thickness and the density,
respectively, ${N_A}$ is the Avogadro constant, ${A_t}$ is the
molecular weight of the target. Assuming implementation of a LiF
target (${\rho}{\simeq}$2.64 g/cm$^{3}$) with a thickness
d of 0.5 cm, $A_t{\simeq}26$, and
$I_i{\simeq}10^{13}$, the expected yield ratio amounts to
$7.4{\times}10^8$/second for exciting the 109.9 keV 1/2$^-$ level
in $^{19}$F.

The accuracy of the  measurement
depends on the details of the experimental set-up (counting rates,
detection solid angles etc.).
According to our estimation we expect to achieve the accuracy
better than 10-20\% for one week measurement,
which exceeds considerably the previous experiments in the
traditional design.
The difficulty for this kind of experimental studies stems from a
high counting ratio of the Compton scattered photons
as background. One method to overcome this problem is to use
a multi-segmented detector in order to greatly reduce the
counting rate of each detector and obtain the necessary total counts
of N${\sim}10^{10}$ as the NRF events. The use of newly developed
lutetium oxyorthosilicate (Lu$_2$SiO$_5$, LSO) and lutetium-yttrium
oxyorthosilicate (Lu$_{2(1-x)}$Y$_{2x}$SiO$_5$, LYSO)
crystals \cite{Qin2005,Chen2004}
with a decay constant of about 40 ns and an energy resolution of 7-10\%
is also  promising for the NRF measurement with a high-counting rate.
Another way is to obtain a photon beam with an
ultra high resolution of ${\Delta}E/E\sim10^{-5} - 10^{-6}$.
In this case, the background photons due to Compton scattering are greatly
reduced, and the ${\gamma}$-ray events due to the NRF process
are relatively enhanced to get a high counting rate
necessary for performing a high-statics PNC measurement.

\begin{ack}

We thank H. Akimune, H. Ejiri, S.~Dat'e,  M. Itoh,
Y.~Ohashi, H. Ohkuma, Y. Sakurai, S. Suzuki, K. Tamura,
H. Toki,
and H. Toyokawa for fruitful discussions.
 One of the authors (A.I.T.) thanks T. Tajima for his
hospitality to stay at SPring-8. This work was strongly
stimulated by a new project to produce a high-intensity MeV
${\gamma}$-rays at SPring-8.

\end{ack}


\end{document}